\documentclass[a4paper,twoside,10pt]{article}
\usepackage{graphicx,publaob}

\usepackage{amsmath,amssymb,amsbsy}
\usepackage{mathrsfs,eucal}

\def\papertype{Contributed paper}
\newcommand{\be}{\begin{equation}}
\newcommand{\ee}{\end{equation}}
\newcommand{\ba}{\begin{eqnarray}}
\newcommand{\ea}{\end{eqnarray}}
\newcommand{\nn}{\nonumber}
\newcommand{\f}{\frac}

\newcommand{\rd}{{\rm d}}
\newcommand{\vp}{\varphi}
\newcommand{\pa}{\partial}

\newcommand{\mm}{\hspace{-0.5mm}}

\begin{document}

\title{Newtonian Aspects of General Relativistic Galaxy Models}

\authors{Aleksandar Raki\'c$^1$ \lowercase{and} Dominik J. Schwarz$^2$}

\address{$^1$ Institut f\"ur Theoretische Physik und Astrophysik, Universit\"at W\"urzburg, D-97074 W\"urzburg, Germany}
\Email{rakic}{astro.uni-wuerzburg}{de}
\address{$^2$ Fakult\"at f\"ur Physik, Universit\"at Bielefeld, D-33501 Bielefeld, Germany}
\Email{dschwarz}{physik.uni-bielefeld}{de}

\markboth{Newtonian Aspects of General Relativistic Galaxy Models}{Aleksandar Raki\'c and Dominik J. Schwarz}

\abstract{Many cosmological observations call for the existence of dark matter. The most direct evidence for dark matter is inferred from the measured flatness of galactic rotation curves. The latter is based on Newtonian gravity. Alternative approaches to the rotation curve problem by means of general relativity have recently been put forward. The class of models of interest is a subset of the axially symmetric and stationary solutions of Einstein's equations with rotating dust. As a step toward the understanding of general relativistic galaxy models, we analyse rigidly as well as non-rigidly rotating (Post-)Newtonian spacetimes. We find that the Newtonian limit of the considered general relativistic galaxy model leads to Post-Newtonian terms in the metric.
}

\section{INTRODUCTION AND MOTIVATION}

Miscellaneous physical observations support the postulation of a new component of matter, non-luminous and only interacting through gravity. The phenomena according to which dark matter is postulated manifest themselves on multiple cosmological scales, ranging from galaxies up to the primordial Universe. The concordance model that enfolds observations of the CMB as well as of the large scale structure currently charges a value of $\Omega_{\rm dm}\simeq0.23$ (WMAP 2008). Herculean efforts are being made in order to find an explanation for the dark matter findings. The attempts range from modifications of the Einsteinian, and therewith of Newtonian, gravity to extensions of the standard model of particle physics that involve new, yet undetected particles that could do the job. But it is important to stress that there is no a priori reason to believe that all of the missing matter problems on many different physical scales have a common explanation.

Here we will be concerned with the dark matter problem on galactic scales. The missing of a Keplerian velocity fall-off ($v(r)\propto 1/\sqrt{r}$) in most of the observed rotation curves represents the main classical evidence for the existence of dark matter in galaxies (Rubin et al.~1980, Binney \& Tremaine 1994). There appears to be far more matter in a galaxy than the luminous part suggests. From the flatness of the rotation curves up to maximally measurable radii, the existence of vast halos of dark matter is concluded. This leads to a cosmological lower bound of $\Omega_{\rm dm} \gtrsim 0.1$ (Amsler et al. 2008). The famous missmatch occurs when comparing the observed rotation curves to Newtonian gravity. But the Newtonian limit appears well justified for a galactic system since the gravitational field strength and the occurring velocities are small.

However, in a recent (Cooperstock \& Tieu 2005-2007) series of works Cooperstock and Tieu (CT) suggest a new approach to the galactic dark matter problem, namely through general relativity (GR). According to these authors the application of full GR to galaxies introduces new effects, cumulating in the bold statement that dark matter can be completely obviated. What may be the motivation for such a try if velocities and fields are small? According to CT, already in a stationary system that is purely gravitationally bound, inherent non-linear terms occur in GR whose magnitudes may not be negligible. Unlike in the Solar System where GR effects are important but minute, CT argue that the constituents of a galaxy, the stars, are not test particles in the field of a central mass but both constitute and follow their mutual common field. This relation does of course not increase the field strength but it shall influence the dynamics of the system in the proper theory. Tangibly, CT fitted flat rotation curves in the GR model and are able to return realistic density profiles using luminous matter only. The resulting integrated masses of CT galaxies lie below the Newtonian values but are larger than the total masses derived from Modified Newtonian Gravity\footnote{For original refs.~see (Milgrom 1983) and for the relativistic extension (Bekenstein 2004).}.

Numerous and severe criticism has been published against the concrete CT model in the meantime. In several works it has been pointed out that there exist pathologies within the energy momentum tensor or various other imponderabilities with the model. For instance Korzy\'nski argues that (A) a proper flat asymptotical limit is not contained in the CT model because of the considered gauge; and that (B) the CT model must be unrealistic because unexpected additional matter sources at $z=0$ can be found (Korzy\'nski 2005): a singular behaviour of the energy-momentum tensor in the plane is revealed because of the existence of a residual non-vanishing Komar mass within a volume that has been shrunk to zero.

Even though the concrete model of CT is problematic the main idea points our attention to GR modelling of galaxies which is a valuable point. This is an important, yet difficult field where less progress has been made than one would expect. There exist only a handful of known axially symmetric solutions in GR and all of them are not viable as a galaxy model, see e.g.~(Islam 1985). The only existing global solution that comes near to this is the one by Neugebauer and Meinel for a disc of rigidly rotating dust (Neugebauer \& Meinel 1994, 1995). This solution was found by the inverse scattering method and its formulation is exceedingly complicated and mostly relevant for numerical modelling. Instead of searching for an exact solution resembling a galaxy, we study the Newtonian limit and the Post-Newtonian contributions to axially symmetric and stationary spacetimes.

\section{AXIALLY SYMMETRIC AND STATIONARY SPACETIMES}

The most general axisymmetric $(\partial_\vp g_{\mu \nu}=0)$ and stationary $(\partial_t g_{\mu \nu}=0)$ spacetime in GR takes the form (Stephani et al. 2003)
\be
 \rd s^2 = e^{-2U} \left[ g_{MN} \rd x^M \rd x^N + W^2 \rd\varphi^2 \right] - e^{2U} ( \rd t + A \rd\varphi)^2 \,,
\label{eq:lpmetric2}
\ee
with notation $(x^0,x^1,x^2,x^3)=(t,r,\vp,z)$ and $M,N$ take values $1$ or $3\,$. It is always possible to go to isotropic coordinates $g_{MN} = e^{2k} \delta_{MN}$ without touching the generality of (\ref{eq:lpmetric2}), c.f.~(Stephani et al.~2003). So we get the final form of the most general axisymmetric and stationary spacetime named after Lewis and Papapetrou (LP)
\be
 \rd s^2 = e^{-2U} \left[ e^{2k}(\rd r^2 + \rd z^2) + W^2 \rd\varphi^2 \right] - e^{2U} ( \rd t + A \rd\varphi)^2 \,,
\label{eq:lpmatricfin}
\ee
with the free metric functions $U,k,W,A$ being solely functions of $r$ and $z\,$.

It is possible to simplify (\ref{eq:lpmatricfin}) a bit more, but only under certain assumptions. We will demonstrate that if and only if the metric function $W$ is harmonic it can be transformed to\footnote{Also $W=1$ is possible then, but this case is of no interest in our case.} $W=r\,$. Consider a complex coordinate transformation $f(r+iz) = W + iV$ introducing an additional potential $V\,$. Then we have from $\rho \equiv W(r,z)$ and $h \equiv V(r,z)$ the differentials $\rd \rho = \f{\pa W}{\pa r} \rd r + \f{\pa W}{\pa z} \rd z\,$ and $\rd h = \f{\pa V}{\pa r} \rd r + \f{\pa V}{\pa z} \rd z\,$. With the coordinates $\rho,h$ being only dummies, introduced for bookkeeping reasons. We insert the transformation into (\ref{eq:lpmatricfin}), written in terms of $\rho,h\,$, and have
\begin{align}
 \rd s^2 &= e^{-2U} \left[ e^{2\tilde{k}}(\rd \rho^2 + \rd h^2) + \rho^2 \rd\varphi^2 \right] - e^{2U} ( \rd t + A \rd\varphi)^2 \longmapsto \rd s^2 = e^{-2U} \Bigg\{ e^{2\tilde{k}} \times \nn \\
 &\times \Bigg[ \left(\f{\pa W}{\pa r}\right)^2 \rd r^2 + \left(\f{\pa W}{\pa z}\right)^2 \rd z^2 + 2 \f{\pa W}{\pa r} \f{\pa W}{\pa z} \rd r \rd z  +  \left(\f{\pa V}{\pa r}\right)^2 \rd r^2 + \left(\f{\pa V}{\pa z}\right)^2 \rd z^2 \nn \\
 &+ 2\f{\pa V}{\pa r} \f{\pa V}{\pa z} \rd r \rd z \Bigg] + W^2 \rd \vp^2 \Bigg\} - e^{2U} ( \rd t + A \rd\varphi)^2 \,.
\label{eq:expltrafo}
\end{align}
Requiring formal invariance as compared to the original metric, we see that the mixing terms should vanish. That is exactly provided by the Cauchy-Riemann equations for $W$ and $V$
\be
 \frac{\partial W}{\partial r} = \frac{\partial V}{\partial z} \quad {\rm and} \quad \frac{\partial W}{\partial z} = - \frac{\partial V}{\partial r} \,.
\ee
Moreover, with the help of the Cauchy-Riemann equations, we see that the coefficients of $\rd r^2$ and $\rd z^2$ can be combined to a positive definite quantity
\be
 \left(\f{\pa W}{\pa r}\right)^2 + \left(\f{\pa V}{\pa r}\right)^2 = \left(\f{\pa W}{\pa z}\right)^2 + \left(\f{\pa V}{\pa z}\right)^2 \equiv \tilde{K} \geq 0 \,,
\ee
such that we can combine $e^{2\tilde{k}}\tilde{K} \equiv e^{2k}$ and so obtain (\ref{eq:lpmatricfin}) via (\ref{eq:expltrafo}). Thus we have shown that it is possible to simplify the general LP form (\ref{eq:lpmatricfin}) by allowing $W = r\,$, which is only possible if the transformation $f$ is analytic, that is $W$ (and also $V$) must be a harmonic function, 
$\Delta^{(2)} W = 0\,$. This condition for $W$ holds for exterior (vacuum) solutions that are stationary and axisymmetric (Islam 1985, Stephani et al.~2003), thus any GR model of a galaxy must approach this class of solutions where the matter density drops to zero. 
Then, we can write down the LP metric in isotropic coordinates 
(or Weyl gauge)
\be
 \rd s^2 = e^{-2U} \left[ e^{2k}(\rd r^2 + \rd z^2) + r^2 \rd\varphi^2 \right] - e^{2U} ( \rd t + A \rd\varphi)^2 \,.
\label{eq:lpweylg}
\ee
The spacetime applied in the CT model is the following 
\be
 \rd s^2 = e^{\nu - w} \left( \rd r^2 + \rd z^2 \right) + r^2 e^{-w} \rd \vp^2 - e^w \left( \rd t + N \rd \vp \right)^2 \,,
\label{eq:ctmetric2}
\ee
together with corotating dust as the matter. Obviously, the CT metric does not belong to the class of the most general stationary and axisymmetric spacetimes; it belongs to the subclass of LP solutions in the Weyl gauge, and it is not obvious that this is justified in the presence of dust.

Now we ask what solution could potentially be a Newtonian counterpart to the CT model (\ref{eq:ctmetric2}). The \lq Newtonian approximation\rq, that is the metric that reproduces Newtonian physics, is given by, see e.g.~(Misner, Thorne \& Wheeler 1973),
\be
  \rd s^2 = -(1 + 2\phi)\rd t^2 + \rd r^2 + r^2 \rd \vp^2 + \rd z^2 \,,
\label{eq:naivenewton}
\ee
where $\phi(r,z)$ is the Newtonian gravitational potential. CT work in a comoving frame, thus we consider the rotation of the Newtonian metric. For simplicity, we start with rigid rotation in (\ref{eq:naivenewton}) via $\varphi = \varphi^\prime - \omega \, t$. The exact result can be brought to the form
\be
 \rd s^2 = (\rd r^2 + \rd z^2) + \frac{1 + 2\phi}{(1 + 2\phi - \omega^2 r^2)} r^2 \rd\varphi^2 - (1 + 2\phi - \omega^2 r^2) \left[\rd t + \frac{r^2 \omega}{(1 + 2\phi - \omega^2 r^2)} \, \rd\varphi \right]^2 \mm\mm.
\label{eq:newstifexa}
\ee
This is the rigidly rotated Newton metric. In this form we can directly compare it with the LP metric in Weyl gauge (\ref{eq:lpweylg}), and we notice a discrepancy at linear order in $\phi\,$, looking at the $\rd \vp^2$ term. Interestingly, the rigidly rotated Newton metric (\ref{eq:newstifexa}) is not consistent with the Weyl subclass of the LP solution (\ref{eq:lpweylg}). Note that (\ref{eq:newstifexa}) is in perfect accordance with the general form of the latter (\ref{eq:lpmatricfin}).

One could now ask whether the situation might be easily cured with the help of a coordinate transformation. We show that this is not possible. Above we have derived the exact conditions under which the general and the isotropic LP metric can be transformed into each other: the function $W$ must be a harmonic function with respect to the two-dimensional Laplacian $\Delta^{(2)} W = 0\,$. In the present case, of the rigidly rotating Newton metric, $W$ is given by
\be
 W = r \sqrt{1 + 2 \phi} \,.
\label{eq:Wdef}
\ee
Expanding to linear order and applying the Laplacian yields
\be
 \Delta^{(2)} W = r \Delta^{(3)}\phi + \phi_{,r} = 4\pi G\rho r + \phi_{,r} \,.
\label{eq:lapl2W}
\ee
Note that we can use the Poisson equation because the potential is Newtonian. After repeating some facts from potential theory we will show that $\Delta^{(2)} W$ in fact does not vanish in general.

Given the general problem of solving the Laplace equation with the appropriate boundary conditions for a disc-like distribution of matter, the solution for the potential can be obtained via separation of variables, c.f.~(Binney \& Tremaine 1994),
\be
 \phi(r,z) = \int_0^\infty S(k) J_0(kr) e^{-k |z|} \rd k \,.
\ee
A given surface mass density $\Sigma(r)$ is then characterised by the according Hankel transform
\be
 S(k) = - 2 \pi G \int_0^\infty J_0(kr) \Sigma(r) r \rd r \,.
\ee
Now we can use these expressions for the evaluation of (\ref{eq:lapl2W}).

Case (A) $z\neq 0$ --- Outside the disc the Newtonian potential fulfils the Laplace equation, such that the expression (\ref{eq:lapl2W}) takes the form
\be
  \Delta^{(2)} W = - \int_0^\infty S(k) J_1(kr) k e^{-k |z|} \rd k \quad {\rm at} \quad z \neq 0 \,,
\ee 
which will not vanish in general. As a simple example we consider the Mestel disc model of a galaxy (Mestel 1963). In the Newtonian Mestel model a flat rotation curve can be reached\footnote{The flat rotation curve in the Mestel model is obtained from the Hankel transform of (\ref{eq:mestelsurfden}), inserted into the formula for the rotation curve: $v^2(r)_{\rm Mes} = r (\pa \phi/\pa r)_{z=0} = 2\pi G\Sigma_0 r_0\,$.}, although for the price of an infinite total mass. The Mestel model is characterised by a surface mass density that falls off inversely with the distance
\be
 \Sigma(r) = \frac{\Sigma_0 r_0}{r} \,.
\label{eq:mestelsurfden}
\ee
In a Mestel galaxy the surface density Hankel-transforms as $ S(k) = - 2 \pi G \Sigma_0 r_0/k $. Using this we
can integrate directly and obtain
\be
 \Delta^{(2)} W = 2 \pi G \Sigma_0 r_0 \left( \f{1}{r} - \f{|z|}{r\sqrt{r^2+z^2}} \right) \quad {\rm at} \quad z \neq 0 \,.
\ee

Case (B) $z = 0$ --- We want to show that (\ref{eq:lapl2W}) is non-zero also here. Let us assume the contrary and see what happens. If we assume that $\Delta^{(2)} W = 0$ was true then equation (\ref{eq:lapl2W}) gives an identity. This we integrate over $z$ for some $\varepsilon > 0$ and then revoke the operation by performing the appropriate limit
\be
 -4 \pi G r \lim_{\varepsilon\to 0} \int_{- \varepsilon}^\varepsilon \delta(z) \Sigma(r) \rd z = \lim_{\varepsilon\to 0} \int_{- \varepsilon}^\varepsilon \int_0^\infty S(k) J_1(kr) k e^{-k|z|} \rd k \rd z \,.
\ee
Because the exponential term on the right hand side acts like a damping factor, the modulus of the integrand will reach its maximum at $z=0$. Thus, as an upper estimate, we can set the integrand of the right hand side to be constant in $z$ and therefore the integration and limit procedure give zero. Then, for all other $z$ the expression will be zero more than ever and we obtain $4 \pi G \Sigma(r) r = 0 \,.$
This will not hold generally for any realistic model, hence producing a contradiction, and therefore $\Delta^{(2)} W(r,z) = 0$ is not true at the surface $z=0$ either.

Interestingly, the pure Newton metric (\ref{eq:naivenewton}) cannot be made compatible with the LP metric in Weyl form and thus is also not compatible with the CT model. 
As all exact vacuum solutions fall into the Weyl class as well, it seems that the Newtonian metric is not appropriate to describe the physics of rotating systems at leading order. We should go one step further and consider the Post-Newtonian (PN) metric
\be
 \rd s^2 = -(1 + 2\phi)\rd t^2 + (1 - 2\psi) (\rd r^2 + r^2 \rd \vp^2 + \rd z^2) \,,
\label{eq:ppnmetric}
\ee
with some additional PN potential $\psi\,$. Sometimes, this metric is referred to as the \lq longitudinal Newton\rq\ metric. The reason might be that the order of magnitude of the coefficient of the spatial part and the order of the Newtonian correction are the same. Nevertheless, conceptually this makes an enormous difference. In classical Newton Gravity there exists no curvature of space, the three-space is always euclidian. This is exactly reflected in the Newton metric (\ref{eq:naivenewton}) and therefore we refer to (\ref{eq:ppnmetric}) as the PN approximation; for an extensive discussion see e.g.~(Misner, Thorne \& Wheeler 1973). 

Now, applying stiff rotation to the PN metric results in (including all orders)
\ba
 \rd s^2 &=& (1 - 2 \psi)(\rd r^2 + \rd z^2) + \frac{(1+2\phi)(1-2\psi)}{[1+2\phi - (1-2\psi)\omega^2 r^2]} r^2\rd\vp^2   \nn \\ 
 &-& [1+2\phi - (1-2\psi)\omega^2 r^2] \left[ \rd t + \frac{ (1-2\psi) \omega r^2}{[1+2\phi - (1-2\psi)\omega^2 r^2]} \, \rd\vp \right]^2 \,.
\label{eq:ppnmetricrotaded}
\ea
For the simplest PN case, i.e.~$\phi=\psi\,$, this is consistent - neglecting terms $\mathcal{O}(\phi^2)$ - with the Weyl gauge (\ref{eq:lpweylg}) of the LP class via
\ba
 & & e^{2 k} = (1-2\phi)[1+2\phi - (1-2\phi)\omega^2 r^2] \,,\quad\; e^{2U} = [1+2\phi - (1-2\phi)\omega^2 r^2] \nn \\
 & & W^2 = r^2 \,,\quad\; A = \frac{ (1-2\phi) \omega r^2}{[1+2\phi - (1-2\phi)\omega^2 r^2]} \,.
\ea
Thus the PN metric after rigid rotation belongs to the class of isotropic (Weyl) axisymmetric and stationary solutions, if $\phi=\psi$, whereas the rotated Newton metric (\ref{eq:newstifexa}) does not allow for that simplification.

Differential rotation is more realistic. So next we relax the condition of rigid rotation and consider transformations $\vp = \vp^\prime - \omega(r) \, t$ with the the PN metric, giving
\ba
 \rd s^2 &=& (1 - 2 \psi)\rd z^2  + (1 - 2 \psi)(1+r^2 \omega_{,r}^2 t^2)\rd r^2  + \frac{(1+2\phi)(1-2\psi)}{[1+2\phi - (1-2\psi)\omega^2 r^2]} r^2\rd\vp^2 \nn \\ 
 &-& [1+2\phi - (1-2\psi)\omega^2 r^2] \left[ \rd t + \frac{ (1 - 2 \psi) \omega r^2}{[1+2\phi - (1-2\psi)\omega^2 r^2]} \rd\vp \right]^2 \nn \\ 
 &+& (1-2\psi)2r^2\omega\omega_{,r} t dr dt - (1-2\psi)2r^2\omega_{,r} t dr
 d\varphi \,.
\label{eq:pnnewtrotr}
\ea
Unfortunately, this metric exhibits direct time dependence in some coefficients. The metric (\ref{eq:pnnewtrotr}) is only reasonable in a strictly local sense. To preserve stationarity we approximate (\ref{eq:pnnewtrotr}) by allowing only for small time intervals or equivalently for small angles of rotation. Then the differentially rotated PN metric (\ref{eq:pnnewtrotr}) takes exactly the same form as (\ref{eq:ppnmetricrotaded}) only with $\omega = \omega(r)\,$, and is thus also, to leading order, compatible with the Weyl gauge of the LP metric for the simplest PN case $\phi=\psi\,$.

Finally, let us note the result for the PN spacetime rotated fully differentially via\footnote{Instead of $\omega$ we use the notation with $\Omega$ to indicate that this can be understood, more generally than angular velocity, as just another allowed metric function.} $\Omega = \Omega(r,z)\,$ which is a lengthy expression we write down in component notation:
\begin{align}
 g_{tt} &= - (1+2\phi) + (1-2\psi) \Omega^2 r^2 \,, \quad g_{tr} = (1-2\psi) tr^2 \Omega \Omega_{,r}t \,,\; g_{t\vp} = -(1-2\psi) r^2 \Omega \nn \\
 g_{tz} &= (1-2\psi)r^2 \Omega \Omega_{,z}t \,,\quad g_{rr} = (1-2\psi) (1+r^2 \Omega_{,r}^2t^2) \,, \quad g_{r\vp} = -(1-2\psi)r^2 \Omega_{,r}t \,, \nn \\
 \quad g_{rz} &= (1-2\psi) r^2 \Omega_{,r} \Omega_{,z} t^2 \,,\quad g_{\vp\vp} = (1-2\psi)r^2 \,, \quad g_{\vp z} = -(1-2\psi) r^2 \Omega_{,z}t \,,\nn \\
 \quad g_{zz} &= (1-2\psi)(1+r^2\Omega_{,z}^2t^2) \,.
\label{eqap:horrend}
\end{align}
Physically relevant however - again we need to meet the constraint of stationarity -, is (\ref{eqap:horrend}) taken as valid in a strictly local sense, that is
\be
 \rd s^2 = [-(1+2\phi)+r^2\Omega^2(1-2\psi)] \rd t^2 + (1-2\psi) ( \rd r^2 + \rd z^2 + r^2 \rd \varphi^2 - 2 r^2 \Omega \rd \varphi \rd t) \,.
\label{eqap:final}
\ee
After some algebra one sees that this is consistent with the Weyl form (\ref{eq:lpweylg}) and can be used as the starting point for a galaxy model in (Post-)Newtonian language. The analysis of the resulting dynamical equations via a 3+1 split is the final step and is going to be published elsewhere.

\section{DISCUSSION}

Motivated by the GR view on the galactic rotation curve problem, we have reviewed the outline of axisymmetric and stationary spacetimes as potential galaxy models. We have seen that a certain proposed model, the CT model, does not employ the most general axisymmetric and stationary spacetime form, the LP form, but belongs to a subclass thereof, the Weyl gauge. It is open in how far this restriction be connected with the difficulties of the CT model.

If there really exist extra terms in a certain general relativistic approach then eventually we should be able to pin down the differences by comparison to well-known Newtonian physics. As a first step we approached the CT metric in a (Post-)Newtonian language. We have shown explicitly that the classical Newton metric after (differential) rotation cannot be brought to Weyl gauge, but is of a more general type. On the other hand, the result of (differential) rotation of the Post-Newtonian spacetime is consistent with the Weyl form in the case of $\phi=\psi\,$. Thus the GR galaxy model considered requires a Post-Newtonian approximation. In a forthcoming publication we are investigating the consequences of these PN terms on galactic dynamics.

\vspace{2mm}
\noindent This work was supported by the DFG under grants GRK 881 and GRK 1147.

\references

Amsler, C. et al. : 2008, \journal{Physics Letters}, \vol{B667}, 1.

Bekenstein, J.~D. : 2004, \journal{Phys.~Rev.~D} \vol{70}, 083509; 2005, \journal{Erratum-ibid.~D} \vol{71} 069901; [astro-ph/0403694].

Binney, J., Tremaine, S. : 1994, \journal{Galactic Dynamics}, Princeton University Press.

Cooperstock, F. I., Tieu, S. : 2005, [astro-ph/0507619].

Cooperstock, F. I., Tieu, S. : 2005, [astro-ph/0512048].

Cooperstock, F. I., Tieu, S. : 2007, \journal{Int. J. Mod. Phys.}, \vol{A22}, 2293; [astro-ph/0610370].

Islam, J. N. : 1985, \journal{Rotating Fields in General Relativity}, Cambridge University Press.

Korzy\'nski, M. : 2005, [astro-ph/0508377].

Mestel, L. : 1963, \journal{Mon.~Not.~Roy.~Astron.~Soc.}, \vol{126}, 553.

Milgrom, M. : 1983, \journal{Astrophys.~J.}, \vol{270}, 365.

Misner, C. W., Thorne, K. S., Wheeler, J. A. : 1973 \journal{Gravitation}, Palgrave Macmillan 1973.

Neugebauer, G., Meinel, R. : 1994, \journal{Phys.~Rev.~Lett.}, \vol{73} 2166.

Neugebauer, G., Meinel, R. : 1995, \journal{Phys.~Rev.~Lett.}, \vol{75} 3046; [gr-qc/0302060].

Rubin, V.~C., Ford, W.~K.~J., Thonnard, N. : 1980, \journal{Astrophys.~J.}, \vol{238}, 471.

Stephani, H., Kramer, D., MacCallum, M.~A.~H., Hoenselaers, C., Herlt, E. : 2003, \journal{Exact solutions of Einstein's field equations}, Cambridge University Press.

WMAP 2008 data products NASA site at: {\tt http://lambda.gsfc.nasa.gov}.

\endreferences

\end{document}